# Combined theory of two- and four- component complete orthonormal sets of spinor wave functions and Slater type spinor orbitals in position, momentum and four-dimensional spaces

I.I.Guseinov

*Department of Physics, Faculty of Arts and Sciences, Onsekiz Mart University, Çanakkale, Turkey*

**Abstract**

By the use of complete orthonormal sets of nonrelativistic scalar orbitals introduced by the author in previous papers the new complete orthonormal basis sets for two-and four-component spinor wave functions, and Slater spinor orbitals useful in the quantum-mechanical description of the spin-1/2 particles by the quasirelativistic and Dirac's relativistic equations are established in position, momentum and four-dimensional spaces. These function sets are expressed through the corresponding nonrelativistic orbitals. The analytical formulas for overlap integrals over four-component relativistic Slater spinor orbitals with the same screening constants in position space are also derived. The relations obtained in this study can be useful in the study of different problems arising in the quasirelativistic and relativistic quantum mechanics when the position, momentum and four dimensional spaces are employed.

## 1. Introduction

It is well known that the eigenfunctions of the Schrödinger equation of the hydrogen-like atom and their extensions to momentum and four-dimensional spaces by Fock [1,2] are not complete unless the continuum is included. In atomic and molecular electronic structure calculations it is quite common to introduce the complete and orthonormal function sets. The examples of complete and orthonormal functions in position space for the particles with spin $s=0$ are so-called Lambda and Coulomb Sturmian orbitals introduced by Hylleraas, Shull and Löwdin in Refs. [3-6]. The relativistic one-electron Coulomb Sturmian wave functions have been developed and discussed by Avery and Antonsen [7].

Vast majority of modern relativistic calculations are being done within numerically more convenient Gaussian type orbitals (GTO) (see [8] and references quoted therein). In the case of the point nuclear model, the GTO do not allow an adequate representation of important properties of the electronic wave function, such as the cusps at the nuclei [9] and the exponential decay at large distances [10]. Therefore, it is desirable to perform the relativistic calculations with the help of exponential type orbitals (ETO), when the point-like nuclear model is used. It should be noted that the straightforward use of relativistic equations in the calculations with finite basis sets, independent for large and small components, could cause numerical instability known as the Brown-Ravenhall disease [11]. A possible solution of this problem is to use "kinetically balanced" basis sets for the small components [12].

The aim of this work is to derive the new complete orthonormal basis sets for two- and four-component exponential type spinor orbitals arising in quasirelativistic and relativistic quantum theory of atomic-molecular and nuclear systems. We notice that the definition of phases in this work for the complete spherical harmonics ($Y_{lm}^* = Y_{l,-m}$) differs from the Condon-Shortley phases [13] by the sign factor $(-1)^m$.

## 2. Nonrelativistic scalar wave functions and Slater orbitals

To construct in position, momentum and four-dimensional spaces the complete orthonormal sets of quasirelativistic and relativistic spinor wave functions, and Slater spinor orbitals for the spin-1/2 particles we shall use the following nonrelativistic scalar orbitals:

$\psi^\alpha$-exponantial type orbitals ($\psi^\alpha - ETO$)

$$\psi_{nlm}^\alpha(\zeta,\vec{r}) = R_{nl}^\alpha(\zeta,r) S_{lm}(\vec{r}/r) \tag{1}$$

$$\overline{\psi}_{nlm}^\alpha(\zeta,\vec{r}) = \overline{R}_{nl}^\alpha(\zeta,r) S_{lm}(\vec{r}/r), \tag{2}$$

$\phi^\alpha$ - momentum space orbitals ($\phi^\alpha - MSO$)

$$\phi_{nlm}^\alpha(\zeta,\vec{k}) = \Pi_{nl}^\alpha(\zeta,k) \tilde{S}_{lm}(\vec{k}/k) \tag{3}$$

$$\overline{\phi}_{nlm}^\alpha(\zeta,\vec{k}) = \overline{\Pi}_{nl}^\alpha(\zeta,k) \tilde{S}_{lm}(\vec{k}/k), \tag{4}$$

$z^\alpha$-hyperspherical harmonics ($z^\alpha - HSH$)

$$z_{nlm}^\alpha(\zeta,\beta\theta\varphi) = P_{nl}^\alpha(\kappa_4) \tilde{T}_{lm}(\kappa_1,\kappa_2,\kappa_3) \tag{5}$$

$$\overline{z}_{nlm}^{\alpha}(\zeta,\beta\theta\varphi) = \overline{P}_{nl}^{\alpha}(\kappa_4)\tilde{T}_{lm}(\kappa_1,\kappa_2,\kappa_3), \tag{6}$$

$\chi$ - Slater type orbitals $(\chi - STO)$

$$\chi_{nlm}(\zeta,\vec{r}) = R_n(\zeta,r)S_{lm}(\vec{r}/r), \tag{7}$$

Slater type u- momentum space orbitals $(u - MSO)$

$$u_{nlm}(\zeta,\vec{k}) = Q_{nl}(\zeta,k)\tilde{S}_{lm}(\vec{k}/k), \tag{8}$$

Slater type $v$ - hyperspherical harmonics $(v - HSH)$

$$v_{nlm}(\zeta,\beta\theta\varphi) = \frac{1}{\zeta^{3/2}}\Gamma_{nl}(\kappa_4)\tilde{T}_{lm}(\kappa_1,\kappa_2,\kappa_3). \tag{9}$$

See Refs. [14-19] for the exact definition of quantities occurring on the right-hand sides of Eqs. (1)-(9). We notice that the quantities $\kappa_1, \kappa_2, \kappa_3$ and $\kappa_4$ occurring in Eqs. (5), (6) and (9) are the Cartesian coordinates on the four-dimensional space. They can be obtained from the components of the momentum vector $\vec{k}$ by the following relations:

$$\begin{aligned}
\kappa_1 &= k_x / \sqrt{\zeta^2 + k^2} = \sin\beta\cos\varphi\sin\theta \\
\kappa_2 &= k_y / \sqrt{\zeta^2 + k^2} = \sin\beta\sin\varphi\sin\theta \\
\kappa_3 &= k_z / \sqrt{\zeta^2 + k^2} = \sin\beta\cos\theta \\
\kappa_4 &= \zeta / \sqrt{\zeta^2 + k^2} = \cos\beta,
\end{aligned} \tag{10}$$

where the angles $\beta,\theta,\varphi$ ($0 \leq \beta \leq \frac{\pi}{2}$, $0 \leq \theta \leq \pi$, $0 \leq \varphi \leq 2\pi$) are spherical coordinates on the four-dimensional unit sphere; $\theta$ and $\varphi$ have the meaning of the usual spherical coordinates in momentum space. The surface element of the four-dimensional sphere is determined by

$$d\Omega(\zeta,\beta\theta\varphi) = \zeta^3 d\Omega. \tag{11}$$

This surface element is connected with the volume element in momentum space by the relation:

$$d^3\vec{k} = dk_x dk_y dk_z = d\Omega(\zeta,\beta\theta\varphi), \tag{12}$$

where

$$d\Omega = d\Omega(1,\beta\theta\varphi) = \frac{\sin^2\beta}{\cos^4\beta} d\beta \sin\theta d\theta d\varphi. \tag{13}$$

In this study, using functions (1)-(9) we introduce in position, momentum and four-dimensional spaces the new complete orthonormal basis sets for two-and four- component spinor wave functions, and Slater spinor orbitals. We notice that the spinor wave functions obtained are complete without the inclusion of the continuum.

**3. Two- and four-component spinor wave functions and Slater type spinor orbitals**

In order to construct the formulas for the two- and four-component spinor wave functions, and Slater spinor orbitals in position, momentum and four-dimensional spaces we use for the spherical spinors the following independent relations:

$$\Omega^l_{jm_j}(\theta,\varphi) = \begin{pmatrix} ta^l_{jm_j}(0)\beta_{m(0)}Y_{lm(0)}(\theta,\varphi) \\ -ta^l_{jm_j}(1)\beta_{m(1)}Y_{lm(1)}(\theta,\varphi) \end{pmatrix} \tag{14}$$

$$\Lambda^{l_t}_{jm_j}(\theta,\varphi) = \begin{pmatrix} -ia^{l_t}_{jm_j}(1)\beta_{m(0)}Y_{l_t m(0)}(\theta,\varphi) \\ -ia^{l_t}_{jm_j}(0)\beta_{m(1)}Y_{l_t m(1)}(\theta,\varphi) \end{pmatrix}. \tag{15}$$

Then, by the use of Eqs. (1)-(9), (14) and (15) one can construct for the two- and four-component of exponential type spinor orbitals in position, momentum and four dimensional spaces the following relations:

complete orthonormal sets of two-component spinor wave functions

$$K^{\alpha l}_{njm_j}(\zeta,\vec{r}) = R^\alpha_{nl}(\zeta,r)\Omega^l_{jm_j}(\theta,\varphi) = \begin{pmatrix} ta^l_{jm_j}(0)\beta_{m(0)}k^\alpha_{nlm(0)}(\zeta,\vec{r}) \\ -ta^l_{jm_j}(1)\beta_{m(1)}k^\alpha_{nlm(1)}(\zeta,\vec{r}) \end{pmatrix} \tag{16a}$$

$$\bar{K}^{\alpha l}_{njm_j}(\zeta,\vec{r}) = \bar{R}^\alpha_{nl}(\zeta,r)\Omega^l_{jm_j}(\theta,\varphi) = \begin{pmatrix} ta^l_{jm_j}(0)\beta_{m(0)}\bar{k}^\alpha_{nlm(0)}(\zeta,\vec{r}) \\ -ta^l_{jm_j}(1)\beta_{m(1)}\bar{k}^\alpha_{nlm(1)}(\zeta,\vec{r}) \end{pmatrix} \tag{16b}$$

$$K^{\alpha l_t}_{njm_j}(\zeta,\vec{r}) = R^\alpha_{nl_t}(\zeta,r)\Lambda^{l_t}_{jm_j}(\theta,\varphi) = \begin{pmatrix} -ia^{l_t}_{jm_j}(1)\beta_{m(0)}k^\alpha_{nl_t m(0)}(\zeta,\vec{r}) \\ -ia^{l_t}_{jm_j}(0)\beta_{m(1)}k^\alpha_{nl_t m(1)}(\zeta,\vec{r}) \end{pmatrix} \tag{17a}$$

$$\bar{K}^{\alpha l_t}_{njm_j}(\zeta,\vec{r}) = \bar{R}^{\alpha}_{nl_t}(\zeta,r)\Lambda^{l_t}_{jm_j}(\theta,\varphi) = \begin{pmatrix} -ia^{l_t}_{jm_j}(1)\beta_{m(0)}\bar{k}^{\alpha}_{nl_t m(0)}(\zeta,\vec{r}) \\ -ia^{l_t}_{jm_j}(0)\beta_{m(1)}\bar{k}^{\alpha}_{nl_t m(1)}(\zeta,\vec{r}) \end{pmatrix}, \qquad (17b)$$

Slater type two-component spinor orbitals

$$K^{l}_{njm_j}(\zeta,\vec{r}) = R_{nl}(\zeta,r)\Omega^{l}_{jm_j}(\theta,\varphi) = \begin{pmatrix} ta^{l}_{jm_j}(0)\beta_{m(0)}k_{nlm(0)}(\zeta,\vec{r}) \\ -ta^{l}_{jm_j}(1)\beta_{m(1)}k_{nlm(1)}(\zeta,\vec{r}) \end{pmatrix} \qquad (18)$$

$$K^{l_t}_{njm_j}(\zeta,\vec{r}) = R_{nl_t}(\zeta,r)\Lambda^{l_t}_{jm_j}(\theta,\varphi) = \begin{pmatrix} -ia^{l_t}_{jm_j}(1)\beta_{m(0)}k_{nl_t m(0)}(\zeta,\vec{r}) \\ -ia^{l_t}_{jm_j}(0)\beta_{m(1)}k_{nl_t m(1)}(\zeta,\vec{r}) \end{pmatrix}, \qquad (19)$$

complete orthonormal sets of four-component spinor wave functions

$$K^{\alpha lt}_{njm_j}(\zeta,\vec{r}) = \frac{1}{\sqrt{2-\delta_{nl_t}}}\begin{pmatrix} K^{\alpha l}_{njm_j}(\zeta,\vec{r}) \\ K^{\alpha l_t}_{njm_j}(\zeta,\vec{r}) \end{pmatrix} \qquad (20a)$$

$$= \frac{1}{\sqrt{2-\delta_{nl_t}}} \begin{pmatrix} ta^{l}_{jm_j}(0)\beta_{m(0)}k^{\alpha}_{nlm(0)}(\zeta,\vec{r}) \\ -ta^{l}_{jm_j}(1)\beta_{m(1)}k^{\alpha}_{nlm(1)}(\zeta,\vec{r}) \\ -ia^{l_t}_{jm_j}(1)\beta_{m(0)}k^{\alpha}_{nl_t m(0)}(\zeta,\vec{r}) \\ -ia^{l_t}_{jm_j}(0)\beta_{m(1)}k^{\alpha}_{nl_t m(1)}(\zeta,\vec{r}) \end{pmatrix} \qquad (20b)$$

$$\bar{K}^{\alpha lt}_{njm_j}(\zeta,\vec{r}) = \frac{1}{\sqrt{2-\delta_{nl_t}}}\begin{pmatrix} \bar{K}^{\alpha l}_{njm_j}(\zeta,\vec{r}) \\ \bar{K}^{\alpha l_t}_{njm_j}(\zeta,\vec{r}) \end{pmatrix} \qquad (21a)$$

$$= \frac{1}{\sqrt{2-\delta_{nl_t}}} \begin{pmatrix} ta^{l}_{jm_j}(0)\beta_{m(0)}\bar{k}^{\alpha}_{nlm(0)}(\zeta,\vec{r}) \\ -ta^{l}_{jm_j}(1)\beta_{m(1)}\bar{k}^{\alpha}_{nlm(1)}(\zeta,\vec{r}) \\ -ia^{l_t}_{jm_j}(1)\beta_{m(0)}\bar{k}^{\alpha}_{nl_t m(0)}(\zeta,\vec{r}) \\ -ia^{l_t}_{jm_j}(0)\beta_{m(1)}\bar{k}^{\alpha}_{nl_t m(1)}(\zeta,\vec{r}) \end{pmatrix}, \qquad (21b)$$

Slater type four-component spinor orbitals

$$K^{lt}_{njm_j}(\zeta,\vec{r}) = \frac{1}{\sqrt{2-\delta_{nl_t}}}\begin{pmatrix} K^{l}_{njm_j}(\zeta,\vec{r}) \\ K^{l_t}_{njm_j}(\zeta,\vec{r}) \end{pmatrix} \qquad (22)$$

$$= \frac{1}{\sqrt{2-\delta_{nl_t}}} \begin{pmatrix} ta^l_{jm_j}(0)\beta_{m(0)}k_{nlm(0)}(\zeta,\vec{r}) \\ -ta^l_{jm_j}(1)\beta_{m(1)}k_{nlm(1)}(\zeta,\vec{r}) \\ -ia^{l_t}_{jm_j}(1)\beta_{m(0)}k_{nl_tm(0)}(\zeta,\vec{r}) \\ -ia^{l_t}_{jm_j}(0)\beta_{m(1)}k_{nl_tm(1)}(\zeta,\vec{r}) \end{pmatrix}. \tag{23}$$

The functions containing in these formulas are defined as

$$k^\alpha_{nlm} = \psi^\alpha_{nlm}(\zeta,\vec{r}), \phi^\alpha_{nlm}(\zeta,\vec{k}), z^\alpha_{nlm}(\zeta,\beta\theta\varphi) \tag{24a}$$

$$\bar{k}^\alpha_{nlm} = \bar{\psi}^\alpha_{nlm}(\zeta,\vec{r}), \bar{\phi}^\alpha_{nlm}(\zeta,\vec{k}), \bar{z}^\alpha_{nlm}(\zeta,\beta\theta\varphi) \tag{24b}$$

$$k_{nlm} = \chi_{nlm}(\zeta,\vec{r}), u_{nlm}(\zeta,\vec{k}), v_{nlm}(\zeta,\beta\theta\varphi) \tag{25}$$

$$K^{\alpha l}_{njm_j} = \Psi^{\alpha l}_{njm_j}(\zeta,\vec{r}), \Phi^{\alpha l}_{njm_j}(\zeta,\vec{k}), Z^{\alpha l}_{njm_j}(\zeta,\beta\theta\varphi) \tag{26a}$$

$$\bar{K}^{\alpha l}_{njm_j} = \bar{\Psi}^{\alpha l}_{njm_j}(\zeta,\vec{r}), \bar{\Phi}^{\alpha l}_{njm_j}(\zeta,\vec{k}), \bar{Z}^{\alpha l}_{njm_j}(\zeta,\beta\theta\varphi) \tag{26b}$$

$$K^l_{njm_j} = X^l_{njm_j}(\zeta,\vec{r}), U^l_{njm_j}(\zeta,\vec{k}), V^l_{njm_j}(\zeta,\beta\theta\varphi) \tag{27}$$

$$K^{\alpha lt}_{njm_j} = \Psi^{\alpha lt}_{njm_j}(\zeta,\vec{r}), \Phi^{\alpha lt}_{njm_j}(\zeta,\vec{k}), Z^{\alpha lt}_{njm_j}(\zeta,\beta\theta\varphi) \tag{28a}$$

$$\bar{K}^{\alpha lt}_{njm_j} = \bar{\Psi}^{\alpha lt}_{njm_j}(\zeta,\vec{r}), \bar{\Phi}^{\alpha lt}_{njm_j}(\zeta,\vec{k}), \bar{Z}^{\alpha lt}_{njm_j}(\zeta,\beta\theta\varphi) \tag{28b}$$

$$K^{lt}_{njm_j} = X^{lt}_{njm_j}(\zeta,\vec{r}), U^{lt}_{njm_j}(\zeta,\vec{k}), V^{lt}_{njm_j}(\zeta,\beta\theta\varphi). \tag{29}$$

The functions $\psi^\alpha_{nlm}, \bar{\psi}^\alpha_{nlm}, \phi^\alpha_{nlm}, \bar{\phi}^\alpha_{nlm}, z^\alpha_{nlm}, \bar{z}^\alpha_{nlm}, \chi_{nlm}, u_{nlm}$ and $v_{nlm}$ occurring in these formulas are determined by Eqs.(1)-(9). It should be noted that the radial parts of orbitals $\psi^\alpha_{nl_tm}, \bar{\psi}^\alpha_{nl_tm}, \phi^\alpha_{nl_tm}, \bar{\phi}^\alpha_{nl_tm}, z^\alpha_{nl_tm}, \bar{z}^\alpha_{nl_tm}, \chi_{nl_tm}, u_{nl_tm}$ and $v_{nl_tm}$ with $l_t = n$ arising in the quasirelativistic and relativistic cases must be equated to zero.

The quantities containing in Eqs.(14)-(23) are determined by

$$1 \leq n < \infty; \ j = l + \frac{1}{2}t \ (t = \pm 1); \ \frac{1}{2} \leq j \leq n - \frac{1}{2}; \ -j \leq m_j \leq j; \ j - \frac{1}{2} \leq l \leq \min(j + \frac{1}{2}, n-1);$$
$$l_t = l + t; \ m(\lambda) = m_j + \lambda - \frac{1}{2}; \ 0 \leq \lambda \leq 1; \ \beta_{m(\lambda)} = (-1)^{[|m(\lambda)|-m(\lambda)]/2} = i^{|m(\lambda)|-m(\lambda)} \tag{30}$$

and

$$a^l_{jm_j}(\lambda) = \left(l\frac{1}{2}m(\lambda)\frac{1}{2}-\lambda\left|l\frac{1}{2}jm_j\right.\right) = t^{1-\lambda}\left[\frac{l+(-1)^\lambda tm_j+\frac{1}{2}}{2l+1}\right]^{1/2}. \tag{31}$$

The coefficients $a^l_{jm_j}(\lambda)$ satisfy the orthonormality relation

$$\sum_{\lambda=0}^{1} a^l_{jm_j}(\lambda)a^l_{j'm_j}(\lambda) = \delta_{jj'}. \tag{32}$$

Using method set out in Ref. [20] it is easy to show that the independent spherical spinors, Eqs. (14) and (15), for the given values of $j$ have the following properties:

$$\vec{\sigma}\hat{\vec{p}}[R^\alpha_{nl}(\zeta,r)\Omega^l_{jm_j}(\theta,\varphi)] = \hbar\left[\frac{dR^\alpha_{nl}(\zeta,r)}{dr}+(1-\kappa)\frac{R^\alpha_{nl}(\zeta,r)}{r}\right]\Lambda^l_{jm_j}(\theta,\varphi), \tag{33}$$

$$\vec{\sigma}\hat{\vec{p}}[R^\alpha_{nl_t}(\zeta,r)\Lambda^{l_t}_{jm_j}(\theta,\varphi)] = -\hbar\left[\frac{dR^\alpha_{nl_t}(\zeta,r)}{dr}+(1+\kappa)\frac{R^\alpha_{nl_t}(\zeta,r)}{r}\right]\Omega^l_{jm_j}(\theta,\varphi), \tag{34}$$

where $\zeta > 0$, $-\infty < \alpha \leq 1$, the three Cartesian components of $\vec{\sigma}$ are the Pauli spin matrices and

$$\kappa = t(j+1/2). \tag{35}$$

Eqs. (33) and (34) are important in the reduction of the Dirac equation to the radial form.

The nonrelativistic scalar, and quasirelativistic and relativistic spinor orbitals satisfy the following orthonormality relations:

$$\int k^{\alpha*}_{nlm}(\zeta,\vec{x})\bar{k}^\alpha_{n'l'm'}(\zeta,\vec{x})d\vec{x} = \delta_{nn'}\delta_{ll'}\delta_{mm'} \tag{36}$$

$$\int k^*_{nlm}(\zeta,\vec{x})k_{n'l'm'}(\zeta,\vec{x})d\vec{x} = \frac{(n+n')!}{\left[(2n)!(2n')!\right]^{1/2}}\delta_{ll'}\delta_{mm'} \tag{37}$$

$$\int K^{\alpha l\dagger}_{njm_j}(\zeta,\vec{x})\bar{K}^{\alpha l'}_{n'j'm'_j}(\zeta,\vec{x})d\vec{x} = \delta_{nn'}\delta_{ll'}\delta_{jj'}\delta_{m_jm'_j} \tag{38}$$

$$\int K^{l\dagger}_{njm_j}(\zeta,\vec{x})K^{l'}_{n'j'm'_j}(\zeta,\vec{x})d\vec{x} = \frac{(n+n')!}{\left[(2n)!(2n')!\right]^{1/2}}\delta_{ll'}\delta_{jj'}\delta_{m_jm'_j} \tag{39}$$

$$\int K^{\alpha l\dagger t}_{njm_j}(\zeta,\vec{x})\bar{K}^{\alpha l't}_{n'j'm'_j}(\zeta,\vec{x})d\vec{x} = \delta_{nn'}\delta_{ll'}\delta_{jj'}\delta_{m_jm'_j} \tag{40}$$

$$\int K^{l^\dagger t}_{njm_j}(\zeta,\vec{x}) K^{l't'}_{n'j'm'_j}(\zeta,\vec{x})d\vec{x} = \frac{(n+n')!}{\left[(2n)!(2n')!\right]^{1/2}} \delta_{ll'}\delta_{jj'}\delta_{m_jm'_j} \tag{41}$$

where

$$\vec{x} = \vec{r}, \vec{k}, \beta\theta\varphi \text{ and } d\vec{x} = d^3\vec{r}, d^3\vec{k}, d\Omega(\zeta,\beta\theta\varphi). \tag{42}$$

As can be seen from the formulas presented in this work, all of the two- and four-component spinor wave functions and Slater type spinor orbitals are expressed through the corresponding nonrelativistic scalar functions defined in position, momentum and four-dimensional spaces. Thus, the expansion and one-range addition theorems obtained in [18] for the $\psi^\alpha$-ETO, $\phi^\alpha$-MSO, $z^\alpha$-HSH and $\chi$-STO can be also used in the case of quasirelativistic and relativistic functions in position, momentum and four-dimensional spaces.

## 4. Evaluation of overlap integrals over Slater type four-component spinor orbitals in position space

Now, we evaluate the two-center overlap integrals over four-component Slater type spinor orbitals with the same screening parameters defined as

$$S^{lt,l't'}_{njm_j,n'j'm'_j}(\vec{G}) = \int X^{lt^\dagger}_{njm_j}(\zeta,\vec{r}) X^{l't'}_{n'j'm'_j}(\zeta,\vec{r}-\vec{R})d^3\vec{r}, \tag{43}$$

where $\vec{r} = \vec{r}_a$, $\vec{r} - \vec{R} = \vec{r}_b$, $\vec{R} = \vec{R}_{ab}$ and $\vec{G} = 2\zeta\vec{R}$. By the use of Eq. (29) for $K^{lt}_{njm_j} \equiv X^{lt}_{njm_j}$ we obtain for integral (43) the following relation:

$$S^{lt,l't'}_{njm_j,n'j'm'_j}(\vec{G}) = \frac{1}{2}\sum_{\lambda=0}^{1}\{tt'a^{l,l'}_{jm_j,j'm'_j}(\lambda)\beta_{m(\lambda),m'(\lambda)}s_{nlm(\lambda),n'l'm'(\lambda)}(\vec{G}) + a^{l_t,l'_t}_{jm_j;j'm'_j}(1-\lambda)\beta_{m(\lambda),m'(\lambda)}s_{nl,m(\lambda);n'l'_t,m'(\lambda)}(\vec{G})\}, \tag{44}$$

where $a^{l,l'}_{jm_j,j'm'_j}(\lambda) = a^l_{jm_j}(\lambda)a^{l'}_{j'm'_j}(\lambda)$, $a^{l_t,l'_t}_{jm_j,j'm'_j}(\lambda) = a^{l_t}_{jm_j}(\lambda)a^{l'_t}_{j'm'_j}(\lambda)$ and $\beta_{m(\lambda),m'(\lambda)} = \beta_{m(\lambda)}\beta_{m'(\lambda)}$. The quantities $s_{nlm(\lambda),n'l'm'(\lambda)}(\vec{G})$ are the overlap integrals over Slater scalar orbitals. They are determined by

$$s_{nlm,n'l'm'}(\vec{G}) = \int \chi^*_{nlm}(\zeta,\vec{r})\chi_{n'l'm'}(\zeta,\vec{r}-\vec{R})d^3\vec{r} \tag{45}$$

It is easy to show that

9$$s_{nlm,n'l'm'}(\vec{G}) = \{[2(n+\alpha)]!/(2n)!\}^{1/2} \sum_{\mu=l+1}^{n+\alpha} \sum_{\mu'=l'+1}^{n'} \frac{1}{(2\mu)^{\alpha}} \overline{\omega}_{n+\alpha,\mu}^{\alpha l} \overline{\omega}_{n'\mu'}^{\alpha l'} s_{\mu lm,\mu'l'm'}^{\alpha}(\vec{G}), \qquad (46)$$

where

$$s_{\mu lm,\mu'l'm'}^{\alpha}(\vec{G}) = \int \overline{\psi}_{\mu lm}^{\alpha*}(\zeta,\vec{r}) \psi_{\mu'l'm'}^{\alpha}(\zeta,\vec{r}-\vec{R}) d^3\vec{r}. \qquad (47)$$

See Ref. [15] for the exact definition of coefficients $\overline{\omega}^{\alpha l}$.

As we see from Eq.(46), the overlap integral of Slater scalar orbitals are expressed in terms of overlap integrals over scalar $\psi^{\alpha}-ETO$. The analytical relations for the evaluation of nonrelativistic overlap integrals over $\psi^{\alpha}-ETO$ were obtained in [18].

The results of calculations for the overlap integrals over Slater type four-component spinor orbitals with the same screening parameters obtained from the complete sets of $\psi^{1}-$, $\psi^{0}-$ and $\psi^{-1}-ETO$ using Mathematica 5.0 international mathematical software are presented in table 1. As we see from the table that the suggested approach guarantees a highly accurate calculation of the overlap integrals over relativistic Slater spinor orbitals.

We notice that the overlap integrals over quasirelativistic and relativistic Slater spinor orbitals with the same screening parameters may be played a significant role in the calculation of arbitrary multicenter integrals arising in position, momentum and four-dimensional spaces when the quantum mechanics is employed for the atomic, molecular and nuclear systems. Thus, the relations for the nonrelativistic overlap integrals over scalar orbitals presented in our previous papers can be used in the evaluation of multicenter integrals over corresponding quasirelativistic and relativistic spinor wave functions and Slater spinor orbitals.

**References**

xyz1. V.A.Fock, Z.Phys., 98 (1935) 145.
2. V.A.Fock, Kgl Norske Videnskab Forh., 31 (1958) 138.
3. E. A. Hylleraas, Z. Phys., 54 (1929) 347.
4. H.Shull, P.O.Löwdin, J. Chem. Phys., 23 (1955) 1362.

**Table1.** The values of overlap integrals over Slater type four-component spinor orbitals obtained from the different complete sets of nonrelativistic $\psi^{\alpha}$-ETO in molecular coordinate system

| n | l | t | j | $m_j$ | n' | l' | t' | j' | $m'_j$ | θ | φ | $G = 2\zeta R$ | $S^{lt,l't'}_{njm_j,n'j'm'_j}(\vec{G})$ | | |
|---|---|---|---|---|---|---|---|---|---|---|---|---|---|---|---|
| | | | | | | | | | | | | | $\alpha = 1$ | $\alpha = 0$ | $\alpha = -1$ |
| 3 | 1 | 1 | 3/2 | 1/2 | 2 | 1 | -1 | 1/2 | 1/2 | 0 | 0 | 10 | -2.4887810380E-01 | -2.4887810380E-01 | -2.4887810380E-01 |
| 6 | 3 | 1 | 7/2 | -1/2 | 5 | 2 | 1 | 5/2 | -1/2 | 0 | 0 | 40 | 7.9055124161E-03 | 7.9055124161E-03 | 7.9055124161E-03 |
| 6 | 4 | -1 | 7/2 | -1/2 | 5 | 3 | -1 | 5/2 | -1/2 | 0 | 0 | 40 | -2.1886363326E-03 | -2.1886363326E-03 | -2.1886363326E-03 |
| 3 | 0 | 1 | 1/2 | 1/2 | 2 | 0 | 1 | 1/2 | 1/2 | π/4 | π/6 | 15 | 5.8206200380E-02 | 5.8206200380E-02 | 5.8206200380E-02 |
| 5 | 3 | -1 | 5/2 | 3/2 | 4 | 2 | -1 | 3/2 | 1/2 | 2π/5 | 5π/3 | 45 | -3.7793727920E-06 | -3.7793727920E-06 | -3.7793727920E-06 |
| 12 | 10 | -1 | 19/2 | 15/2 | 10 | 6 | 1 | 13/2 | 9/2 | π/6 | π/4 | 90 | -6.1611871954E-07 | -6.1611871954E-07 | -6.1611871954E-07 |